\documentclass[preprintnumbers]{revtex4}
\UseRawInputEncoding
\usepackage{amssymb}
\usepackage{amsmath}
\usepackage{graphicx}
\usepackage{dcolumn}
\usepackage{bm}
\usepackage{subfigure}
\usepackage{color}
 \usepackage{CJKutf8}
\numberwithin{equation}{section}
\numberwithin{figure}{section}

\begin{document}

\title{Schottky Anomaly of Reissner-Nordstr\"{o}m-de Sitter spacetime}
\author{Hai-Long Zhen,$^{1,2}$ Yu-Bo Ma,$^{1,2}$ Huai-Fan Li,$^{1,2}$ Li-Chun Zhang,$^{1,2}$ Yun-Zhi Du,$^{1,2}$}
\thanks{\emph{e-mail: duyzh13@lzu.edu.cn}(corresponding author)}

\address{$^1$Department of Physics, Shanxi Datong University, Datong 037009, China\\
$^2$Institute of Theoretical Physics, Shanxi Datong University, Datong, 037009, China\\}

\begin{abstract}

In the extended thermodynamics of black holes, there exists a thermodynamical pressure whose dual thermodynamical quantity is volume. Extensive studies have been conducted on the phase structure of numerous black holes, which have demonstrated striking similarities to the phase structure of various ordinary matter systems. From the comparison of the thermodynamic properties between spherically symmetric AdS black holes and ordinary thermodynamic systems we known that the isovolumetric heat capacity of the former is zero, whereas that of the latter is non-zero. It is a subject of interest for the intrinsic reason for this discrepancy. For the Reissner-N\"{o}rdstrom-de Sitter (RN-dS) spacetime with the coexistence of the black hole and cosmological horizons the effective thermodynamic quantities as well as the interaction between two horizons are presented. The heat capacity in the Reissner-N\"{o}rdstrom-de Sitter (RN-dS) spacetime is then investigated, and it is demonstrated that the behavior of the heat capacity in the RN-dS spacetime is analogous to that of Schottky specific heat. Treating two horizons in the RN-dS spacetime as two distinct energy levels in a two-energy-level system we investigate the thermodynamic properties in the RN-dS spacetime with the method of studying the thermodynamic properties in an ordinary two-energy system, thereby elucidating the intrinsic reasons for the occurrence of Schottky specific heat in the RN-dS spacetime. The heat capacity observed in the RN-dS spacetime is not only consistent with that of the Schottky specific heat described by the effective thermodynamic quantities in the RN-dS spacetime, but also with that of an ordinary two-energy-level system. These results not only reveal the quantum properties of the RN-dS spacetime, but also provide a new avenue for further in-depth study of the quantum properties of black holes and dS spacetime.
\end{abstract}

\maketitle

\section{Introduction}

On 10 April 2019, the Event Horizon Telescope (EHT) research team released the first photograph of the supermassive black hole at the centre of the galaxy M87 \cite{M871,M872}. The morphology of the photograph is consistent with the Kerr black hole model predicted by general relativity. This observation provides compelling evidence for the presence of a supermassive black hole in the central engine of active galactic nucleus. Black hole is not only a strong gravitational system, but also a thermodynamic system. As early as the 1970s, physicists such as Hawking and Bekenstein established the four laws of black hole thermodynamics \cite{HSW,BJD}. The investigation of the thermodynamic properties of black holes not only contributes to a deeper understanding of their nature, but also provides insights into their internal microscopic state.

In order to investigate the microstructure of black holes as well as the formation and evolution of black holes in the Universe, the study of the thermodynamic properties of AdS and dS spacetime has attracted significant attention. Recently, in the context of the $n$-dimensional AdS spacetime, a potential correlation between the cosmological constant, $\Lambda$, and thermodynamical pressure, $P$, has been proposed. This hypothesis, which can be found in  \cite{Kubiznak2012,Cai2013}, has attracted significant attention. This relation is defined as $P=\frac{n(n-1)}{16\pi l^2},~\Lambda=\frac{n(n-1)}{2l^2}$,
where the thermodynamical volume, $V$, is the conjugate quantity to the pressure as $V=\left(\frac{\partial M}{\partial P}\right)_{S,Q_i,J_k}$.
Based on these considerations, the thermodynamic properties of AdS black holes have been studied from different perspectives. First, by treating the AdS black hole as an ordinary thermodynamic system, the phase transition of AdS black hole has been studied in comparison with the phase transition of the ordinary system. It has been demonstrated that the phase transition of the AdS black hole is analogous to that of the ordinary van der Waals system \cite{Cai2016,JLZhang2015,Gunasekaran2012,Frassino2014,Altamirano2014,JYYang2023,JWu2023,SWWei2303,Durgut2023,JWu20231,YXiao2024,MTW2023,RZhao2013,Hendi2017,Hendi2023,Momennia2021,Hendi2021,Laassiri2401,XQLi2023,Kruglov2022,Kruglov20221,YZDu2022,YZDu2023,ZYGao2022,Ghaffari2312}. Secondly, the John-Thomson coefficients of AdS black holes have been investigated. The effects of different parameters and dimensions of AdS black holes on the John-Thomson coefficient have been explored \cite{MYZhang2401,DJGogoi2023,Yasir2024,Barrientos2022,Sadeghi2402}. Recently, the study of the topological properties of AdS black holes has attracted interest. The properties of black holes at the phase transition point are classified through the study of topological numbers \cite{Hunga2023,Gogoi2023,Fairoos2304,DWu2402,Yerra2304}. The microstructure of black holes has been one of the most significant theoretical problems. Notable advances have been made in exploring the microstructure of black holes using various methods \cite{SWWei2015,SWWei20151,SWWei2019,DYChen2023,Sokoliuk2024,Ruppeiner2023,Masood2023,GRLi2022,XYGuo2019}. Nevertheless, the microstructure of black holes remains a mystery. In order to gain insight into the quantum properties and microstructure of black holes, it is necessary to have a comprehensive understanding of the thermodynamic properties of black holes. Consequently, the discussion of the various thermodynamic properties of black holes remains a significant topic in contemporary theoretical physics.

During the early inflationary epoch, our universe is regarded as a quasi-de Sitter spacetime. If the cosmological constant corresponds to dark energy, our universe will evolve into a new de Sitter phase \cite{RGCai2002}. To construct the entire history of the evolution of the Universe and to identify the intrinsic cause of the accelerated expansion of the Universe, it is necessary to have a clear understanding of the classical, quantum and thermodynamic properties of the de Sitter spacetime. Consequently, the investigation of the thermodynamic characteristics of de Sitter spacetime has attracted considerable attention \cite{Marks2107,Sekiwa2006,Urano2009,Mbarek2019,Kubiznak2016,Simovic2019,Haroon2020,Simovic2008,Dolan2013,Bhattacharya2016,McInerney2016,Kanti2017,LCZhang2016,LCZhang2019,YBMa2020,Ko2312,Chakrabhavi2311,Anderson2022,Dinsmore2020,YBDu2303}. It is well established that in de Sitter spacetime, the black hole horizon and the cosmological horizon coexist if the spacetime parameters satisfy certain conditions. In this region, the requirement of thermodynamic equilibrium stability in spacetime is not satisfied due to the existence of two horizon interfaces with different radiation temperatures. Consequently, in order to discuss the thermodynamic properties of this region, it is necessary to establish a thermodynamic system that satisfies the thermodynamic equilibrium stability requirement. Because the thermodynamic system must satisfy the universal relation for thermodynamic systems, which is the first law of thermodynamics. First, the state parameters of the effective thermodynamic system established in dS spacetime satisfy the first law of thermodynamics. Secondly, the effective thermodynamic quantities satisfy the boundary conditions when the values of the state parameter in spacetime are those of the boundary of the coexistence region between the two horizon interfaces. Consequently, the effective thermodynamic system, based on the aforementioned conditions, provides a more comprehensive representation for the thermodynamic properties in spacetime.

A differential equation for the entropy of an effective thermodynamic system is presented, based on the consideration of the interaction between the two horizon interfaces in dS spacetime. This is for the state parameter satisfying the universal first law of thermodynamics. The solution for the entropy of the effective thermodynamic system is obtained by the condition satisfied by the effective thermodynamic system at the boundary. Subsequently, the effective temperature and effective pressure of the effective thermodynamic system are then determined, and the equation of state of the effective thermodynamic system satisfying the boundary conditions is found. This provides a foundation for further in-depth study of the thermodynamic effect in dS spacetime. The study reveals that when the thermodynamic properties are described by the effective thermodynamic quantities in RN-dS spacetime, the behavior of the heat capacity in RN-dS spacetime is analogous to that of the Schottky specific heat. This is highly consistent with the behavior of the heat capacity when the two horizons in RN-dS spacetime are treated as a two-energy-level system. This conclusion contributes to a deeper understanding of dS spacetime, which can be used to employed to investigate the state of motion of microscopic particles within black holes and to simulate the evolution of the Universe. Furthermore, this discovery offers a novel avenue for investigating the intrinsic causes of the accelerated expansion of the Universe.

The paper is arranged as follows: in Sec. \ref{two}, the conditions for the presence of the black hole horizon and the cosmological horizon in RN-dS spacetime, along with the effects of parameters in spacetime on the existence of spacetime with the black hole horizon and the cosmological horizon are discussed. Furthermore, the interval of the value of the position ratio, $x$, between the two horizons in spacetime is determined. In Sec. \ref{three}, the effective thermodynamic quantities in the RN-dS spacetime satisfying the boundary conditions are presented. The behavior of the effective temperature, $T_{\it{eff}}$, the radiation temperature of the black hole horizon, $T_+$, and the radiation temperature of the cosmological horizon, $T_c$, with respect to the position ratio between the two horizons, $x$, in spacetime is analysed. The Smarr relation is expressed by treating effective quantities as state parameters. In Sec. \ref{four}, the behavior of the heat capacity, $C_{Q,l}$, is investigated in the RN-dS spacetime, under the conditions where both the charge, $Q$, and the cosmological constant, $l$, are fixed. The results demonstrate that the behavior of the heat capacity, $C_{Q,l}$, in the RN-dS spacetime with the effective temperature, $T_{\it{eff}}$, and the position ratio, $x$, between the two horizons, exhibits Schottky peaks. This is similar to the behavior of the heat capacity, $\hat{C}_{Q,l}$, with the effective temperature, $T_{\it{eff}}$, and the position ratio between the two horizons, $x$, in a two-level system treating the black hole horizon and the cosmological horizon as two distinct energy levels within an effective thermodynamic system. It indicates that the Schottky peaks in RN-dS spacetime is determined by a two-energy-level system comprising two distinct horizons. Finally, a brief summary is presented in Sec. \ref{five}
\section{RN-dS spacetime}\label{two}

The RN-dS black hole is a static solution of the Einstein-Maxwell equations, and it is also the one of incorporating an additional charge to the Schwarzchild-dS case (see ref. \cite{Ko2312}). The metric of the RN-dS black hole in a four-dimensional dS spacetime is
\begin{align}\label{2.1}
ds^2&=-g(r)dt^2+g^{-1}dr^2+r^2d\Omega_{2}^{2}
\end{align}
with the horizon function
\begin{align}\label{2.2}
g(r)&=1-\frac{2M}{r}+\frac{Q^2}{r^2}-\frac{r^2}{l^2}.
\end{align}
$M$ and $Q$ represent the mass and charge of the black hole, respectively. The curvature radius of dS space is denoted by $l$. The black hole and cosmological horizons ($r_{+}$ and $r_{c}$) are satisfied the equation $g(r_{+,c})=0$. Substituting eq. (\ref{2.2}) into this relation we get
\begin{align}\label{2.3}
M&=\frac{r}{2}\left(1+\frac{Q^2}{r^2}-\frac{r^2}{l^2}\right).
\end{align}
The $M-r$ curve is then drawn from eq. (\ref{2.3}).
\begin{figure}[htb]
\centering
\includegraphics[width=8.5cm,height=4.5cm]{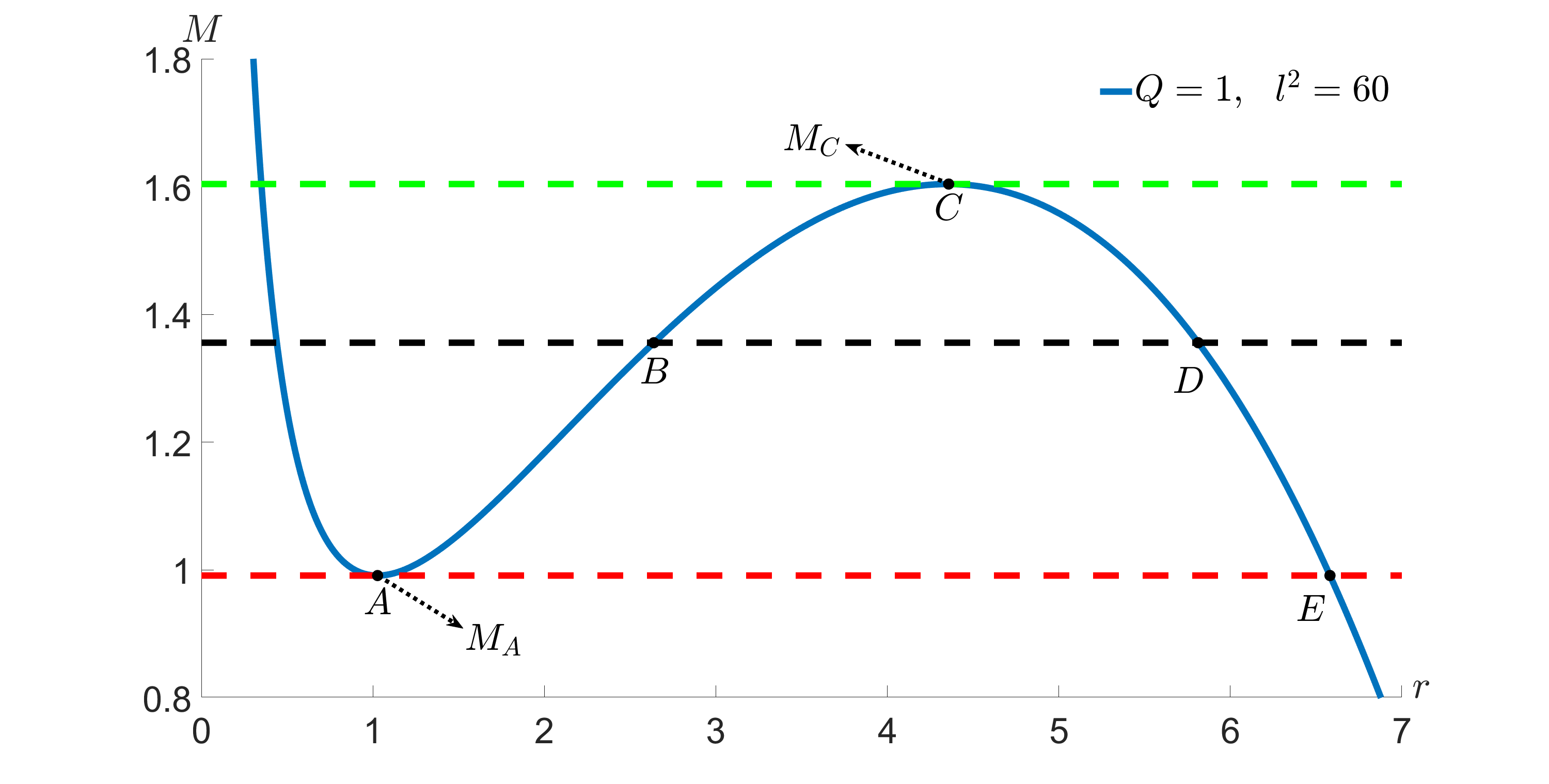}
\vskip -4mm \caption{The $M-r$ curve}\label{fig2.1}
\end{figure}

From Fig. \ref{fig2.1}, it can be observed that the $M-r$ curve demonstrates that the RN-dS spacetime possesses an inner black hole horizon $r_-$, a black hole horizon $r_+$, and a cosmological horizon $r_c$ in case of the parameters $Q$=1, $l^2$=60. The corresponding mass parameter of the RN-dS spacetime with the coexistence of the black hole and cosmological horizons satisfies the expression $M_A\leq{M}\leq{M_c}$. At the point of local maxima designated as $C$, the black hole horizon and cosmological horizon coincide. This point corresponds to the maximum energy, $M=M_c$, where both the black hole and cosmological horizons exist. At the point of the local minimum, designated as $A$, the inner and outer horizons of the black hole coincide. This point corresponds to the minimum energy, $M=M_A$, where both the black hole and inner black hole horizon exist in spacetime. In the RN-dS spacetime, a black hole does not exist when $M>M_c$ or $M<M_A$ \cite{Nam2018}. The Nariai radius and mass had been given in ref. \cite{Nam2018}
\begin{align}\label{2.4}
\nonumber
r_N^2&=\frac{l^2}{6}\left(1+\sqrt{1-\frac{12Q^2}{l^2}}\right),
\\
M_N&=\frac{l}{\sqrt{6}\left(1+\sqrt{1-\frac{12Q^2}{l^2}}\right)}\left(\frac{1}{3}\left(1+\sqrt{1-\frac{12Q^2}{l^2}}\right)+\frac{4Q^2}{l^2}\right). \end{align}
When $Q^2\ll{l^2}$, $M_A\rightarrow0$ and ${M_C}\rightarrow{\frac{1}{3\sqrt{3}}}$. From Eq. (\ref{2.4}) we can see that the charge $Q^2$ and the cosmological constant $l^2$ should satisfy the condition $\frac{12Q^2}{l^2}\leq{1}$. When the local minimum and local maximum points of the black hole mass curve merge into an inflection point, the three horizons coincide. In this case, the black hole is referred to as the ultracold black hole, whose horizon radius and mass are as follows
\begin{align}\label{2.5}
r_{AC}^{4}&=\frac{Q^{2}l^{2}}{3},~~r_{AC}^{4}=4Q^{4},~~M_{AC}=\frac{4Q}{3\sqrt{2}}.
\end{align}
From Eq. (\ref{2.4}), it can be seen that when $Q$=0, $r_{A,C}^{2}=\frac{l^2}{3}$ there is no coexistent region of two horizons. By treating the black hole and the cosmological horizon as independent thermodynamic systems, it can be demonstrated that the thermodynamic quantities on the two horizons each satisfy the first law of thermodynamics \cite{RGCai2002,Dolan2013,Nam2018}.

For the black hole horizon, the associated thermodynamic quantities are
\begin{align}\label{2.7}
T_{+}&=\frac{g'(r_+)}{4\pi}=\frac{1}{4{\pi}r_+}\left(1-\frac{3r_{+}^{2}}{l^2}-\frac{Q^2}{r_{+}^{2}}\right),~~S_+={\pi}r_{+}^{2},~~\phi_{+}=\frac{Q}{r_+}.
\end{align}
which satisfy the following expression
\begin{align}\label{2.6}
dM&=T_{+}dS_{+}+V_{+}dP+\phi_{+}dQ.
\end{align}
The corresponding thermodynamic quantities associated with the cosmological horizon are
\begin{align}{\label{2.9}}
T_{c}&=-\frac{g'(r_c)}{4\pi}=-\frac{1}{4{\pi}r_c}\left(1-\frac{3r_{c}^{2}}{l^2}-\frac{Q^2}{r_{c}^{2}}\right),~~S_c={\pi}r_{c}^{2},~~\phi_{c}=\frac{Q}{r_c}.
\end{align}
and these quantities also satisfy the first law of thermodynamic
\begin{align}\label{2.8}
dM&=-T_{c}dS_{c}+V_{c}dP+\phi_{c}dQ.
\end{align}

From the expression $g(r_{+,c}) = 0$ we can derive
\begin{align}{\label{2.10}}
M&=\frac{r_{c}x(1+x)}{2(1+x+x^2)}+\frac{Q^2(1+x)(1+x^2)}{2r_{c}x(1+x+x^2)},~~\frac{1}{l^2}=\frac{1}{r_{c}^{2}(1+x+x^2)}-\frac{Q^2}{r_{c}^{4}(1+x+x^2)},
\end{align}
where $x=r_+/r_c$, which denotes the ratio of the position of the black hole horizon to that of the cosmological horizon. By solving Eq. (\ref{2.10}) we can obtain two solutions of $r_{c}^2$: a '+' sign before one radical (a larger root) and a '-' sign before the other (a smaller one), which corresponds to the point $C$ when $x\rightarrow1$. Consequently, only the '$+$' sign root corresponding to the larger root is the physical solution and it has the following form
\begin{align}{\label{2.11}}
r_c^2&=\frac{l^2}{6}\left(1+\sqrt{1-\frac{12Q^2}{l^2}}\right).
\end{align}
As $l^2={\alpha}12Q^2$ (with ${\alpha}\geq{1}$ being constant), from Eq. (\ref{2.10}) we can get
\begin{align}{\label{2.12}}
r_c^2&=\frac{6{\alpha}Q^2}{(1+x+x^2)}\left(1\pm\sqrt{1-\frac{(1+x+x^2)}{3x\alpha}}\right),~~\frac{1}{r_c^2}=\frac{x}{2Q^2}\left(1-\sqrt{1-\frac{(1+x+x^2)}{3x\alpha}}\right),
\end{align}
the position of the ultracold black hole horizon at $\alpha=1$ and $x\rightarrow{1}$ is given by Eq. (\ref{2.12}).

Substituting Eq. (\ref{2.10}) into Eqs. (\ref{2.7}) and (\ref{2.9}) yields the radiation temperatures of the black hole horizon and the cosmological horizon, respectively, as
\begin{align}{\label{2.13}}
T_{+}&=\frac{(1-x)}{4{\pi}r_{c}x}\left(\frac{(1+2x)}{(1+x+x^2)}-\frac{Q^{2}(1+2x+3x^2)}{r_{c}^{2}x^{2}(1+x+x^2)}\right),
\end{align}
\begin{align}{\label{2.14}}
T_{c}&=\frac{(1-x)}{4{\pi}r_{c}}\left(\frac{(2+x)}{(1+x+x^2)}-\frac{Q^{2}(3+2x+x^2)}{r_{c}^{2}x(1+x+x^2)}\right).
\end{align}
From Fig. \ref{fig2.1}, it can be observed that the black hole horizon and the cosmological horizon appear simultaneously in spacetime in the case of the electric charge, $Q$, the cosmological constant , $l$, fixed, as well as the energy in spacetime meets the constraints $M_A\leq{M}\leq{M_C}$. The energy, $M$, the electric charge, $Q$, and the cosmological constant, $l$, corresponding to the two points $B$ and $D$ are identical. Nevertheless, in general, the radiation temperature of the black hole horizon ($B$ point), $T_+$, is not equal to the radiation temperature of the cosmological horizon ($D$ point), $T_c$. In this region, two thermodynamic subsystems with different temperatures emerge in spacetime, corresponding to the thermodynamic system on the black hole horizon and the cosmological horizon, respectively. Consequently, under the same parameters, the spacetime in this region is in a state of thermodynamic equilibrium that is unstable. At point $A$, the radiation temperature of the black hole is zero, which also marks the end of the coexistence region of two different radiation temperatures in spacetime. Point $E$, which corresponds to point $A$, represents the endpoint of spacetime from two thermodynamic subsystems to the thermodynamic system only including the cosmological horizon, as well as the endpoint of spacetime with the two thermodynamic subsystems. The RN-dS spacetime exhibits disparate thermodynamic characteristics across distinct regions. Of particular interest is the thermodynamic behavior in the coexistence region, where two disparate temperatures, $T_+$ and $T_c$, represent a topic of great theoretical interest \cite{RGCai2002,Sekiwa2006,Urano2009,Mbarek2019,Kubiznak2016,Simovic2019,Haroon2020,Simovic2008,Dolan2013,Bhattacharya2016,McInerney2016,Kanti2017,LCZhang2016,LCZhang2019,YBMa2020,Ko2312,Chakrabhavi2311,Anderson2022,Dinsmore2020,YBDu2303,Nam2018}.

In recent years, a search has been conducted for the state quantities of thermodynamic systems in spacetime in the region of coexistence of two subthermodynamic systems under various conditions. In \cite{Sekiwa2006,McInerney2016}, the sum of the entropies corresponding to the two horizons is defined as the total entropy of spacetime. The effective temperature of the thermodynamic system in spacetime is then obtained by satisfying the first law of thermodynamics, which requires that the system in question be in equilibrium. This concept paves the way for the investigation of the thermodynamic properties of non-equilibrium states in spacetime. However, the interaction between two horizons is not considered, and the conclusions obtained are therefore not comprehensive. In \cite{LCZhang2016,YBMa2020}, the effective temperature and interaction entropy of spacetime are presented based on the interaction between the two horizons. This provides a foundation for the investigation of the thermodynamic properties of dS spacetime and the entropic force between the two horizons. It is crucial to acknowledge that this approach does not account for the condition that when the spacetime is at the endpoint of the two subthermodynamic systems, $E$, only the temperature of the cosmological horizon, $T_c$, exists in spacetime, with the radiation temperature of the black hole horizon, $T_+$, being zero. This paper presents an extension of the study of the thermodynamic properties of dS spacetime, as previously outlined in \cite{LCZhang2016,YBMa2020}, to encompass the effective thermodynamic quantities of the RN-dS spacetime under various known boundary conditions.
\section{The effective thermodynamic quantities in RN-dS spacetime} \label{three}

The analyses presented in Sec. \ref{two} demonstrate that in order to establish a global thermodynamic system in a region with two subsystems in spacetime, certain boundary conditions must be satisfied by the system and be universal. The thermodynamic properties in RN-dS spacetime, which exhibits a black hole horizon and a cosmological horizon, are being considered. It is therefore of interest to consider the interval $M_A\leq{M}\leq{M_C}$. The exact values of $M_A$ and $M_C$ are related to the spacetime parameters, such as charge, $Q$, and the cosmological constant, $l$, which are given by Eq. (\ref{2.10}).
Considering the connection between the black hole horizon and the cosmological horizon, the effective thermodynamic quantities and the corresponding first law of black hole thermodynamics are derived as follow \cite{LCZhang2016,YBMa2020}.
\begin{align}{\label{3.1}}
dM&=T_{\it{eff}}dS-P_{\it{eff}}dV+\phi_{\it{eff}}dQ.
\end{align}
Here the thermodynamic volume is that between the black hole horizon and the cosmological horizon, namely \cite{Sekiwa2006,Dolan2013,McInerney2016}
\begin{align}{\label{3.2}}
V&=V_{c}-V_{+}.
\end{align}
From Eq. (\ref{2.13}), it can be seen when the potential, denoted by $\phi_A$, on the black hole horizon at point $A$ satisfies the following expression
\begin{align}{\label{3.3}}
\frac{Q^2}{r_{A}^{2}}&=\frac{1+2x}{1+2x+3x^2}.
\end{align}
The radiation temperature of the black hole horizon, $T_+$=0. The radiation temperature of the cosmological horizon, which corresponds to the point $E$, is
\begin{align}{\label{3.4}}
T_E=\frac{(1-x)^2(1+x)}{2{\pi}r_{c}(1+2x+3x^2)}.
\end{align}
When the radiation temperature of black hole is taken as that at point $A$, i.e., the black hole temperature is zero, under the same parameters the effective temperature of RN-dS spacetime should be $T_E$, at the same time the potential on the black hole horizon in the RN-dS spacetime satisfies Eq. (\ref{3.3}). In this context, the effective temperature in spacetime, $T_{\it{eff}}$, can be derived from Eqs. (\ref{3.1}), (\ref{3.2}) and the boundary condition (\ref{3.4}) as follow
\begin{align}{\label{3.5}}
T_{\it{eff}}&=\frac{1-x}{4{\pi}r_{c}x^5}\left\{\left[\left(1+x\right)\left(1+x^3\right)-2x^2\right]-\frac{Q^2}{r_{c}^{2}x^2}\left[\left(1+x+x^2\right)\left(1+x^4\right)-2x^3\right]\right\},
\end{align}
the effective pressure in spacetime, $P_{\it{eff}}$, is
\begin{align}{\label{3.6}}
P_{\it{eff}}&=-\frac{(1-x)}{16{\pi}r_{c}^{2}x^{5}}\left[F'(x)\left(x(1+x)-\frac{Q^2(1+x+x^2+x^3)}{r_{c}^{2}x}\right)-\frac{2F(x)}{(1+x+x^2)}\left((1+2x)-\frac{Q^2(1+2x+3x^2)}{r_{c}^{2}x^2}\right)\right]\notag
\\
&=-\frac{(1-x)}{8{\pi}r_{c}^{2}x^{5}(1-x^3)}\left[\frac{x^6}{(1-x^3)}\left((1+x)-\frac{Q^2(1+x+x^2+x^3)}{r_{c}^{2}x^2}\right)\notag\right.
\\
&\left.-F(x)\left(\left[(1+x)(1+x^3)-2x^2\right]-\frac{Q^2\left[(1+x+x^2)(1+x^4)-2x^3\right]}{r_{c}^{2}x^2}\right)\right]
\end{align}
with
\begin{align}{\label{3.7}}
F(x)&=\frac{8}{5}(1-x^3)^{2/3}+\frac{2}{5(1-x^3)}-1=1+x^2+f_0(x),\\ \nonumber
f_0(x)&=\frac{8}{5}(1-x^3)^{2/3}-\frac{8+5x^2-10x^3-5x^5}{5(1-x^3)}.
\end{align}
The thermodynamic volume, $V$, entropy, $S$ and potential, $\phi_{\it{eff}}$, of the effective thermodynamic system are, respectively
\begin{align}\label{3.8}
V=\frac{4\pi}{c}r_{c}^{3}(1-x^3),~~S={\pi}r_{c}^{2}F(x),~~{\phi}_{\it{eff}}=\frac{Q(1+x)(1+x^2)}{r_{c}x(1+x+x^2)}.
\end{align}
Note that the effective thermodynamic temperature, pressure, and entropy of RN-dS spacetime are different from that shown in ref. \cite{YZDu2022}. This is due to the fact that we have considered different boundary conditions although the thermodynamical volume has the same form. From Eqs. (\ref{2.10}) and (\ref{3.3}), the minimum value, $x=x_{\it{min}}$, of the ratio, $x$, between the two horizon positions in spacetime, given the charge, $Q$, and the cosmological constant, $l$, in spacetime, is given by
\begin{align}\label{3.9}
\frac{6{\alpha}}{(1+x+x^2)}\left(1\pm\sqrt{1-\frac{(1+x+x^2)}{3x\alpha}}\right)&=\frac{1+2x+3x^2}{x^2(1+2x)},
\end{align}
i.e. $(1+2x=3x^2)^2-12{\alpha}x^2(1+2x)=0$.
\begin{figure}[htb]
\centering
\includegraphics[width=8.5cm,height=4.5cm]{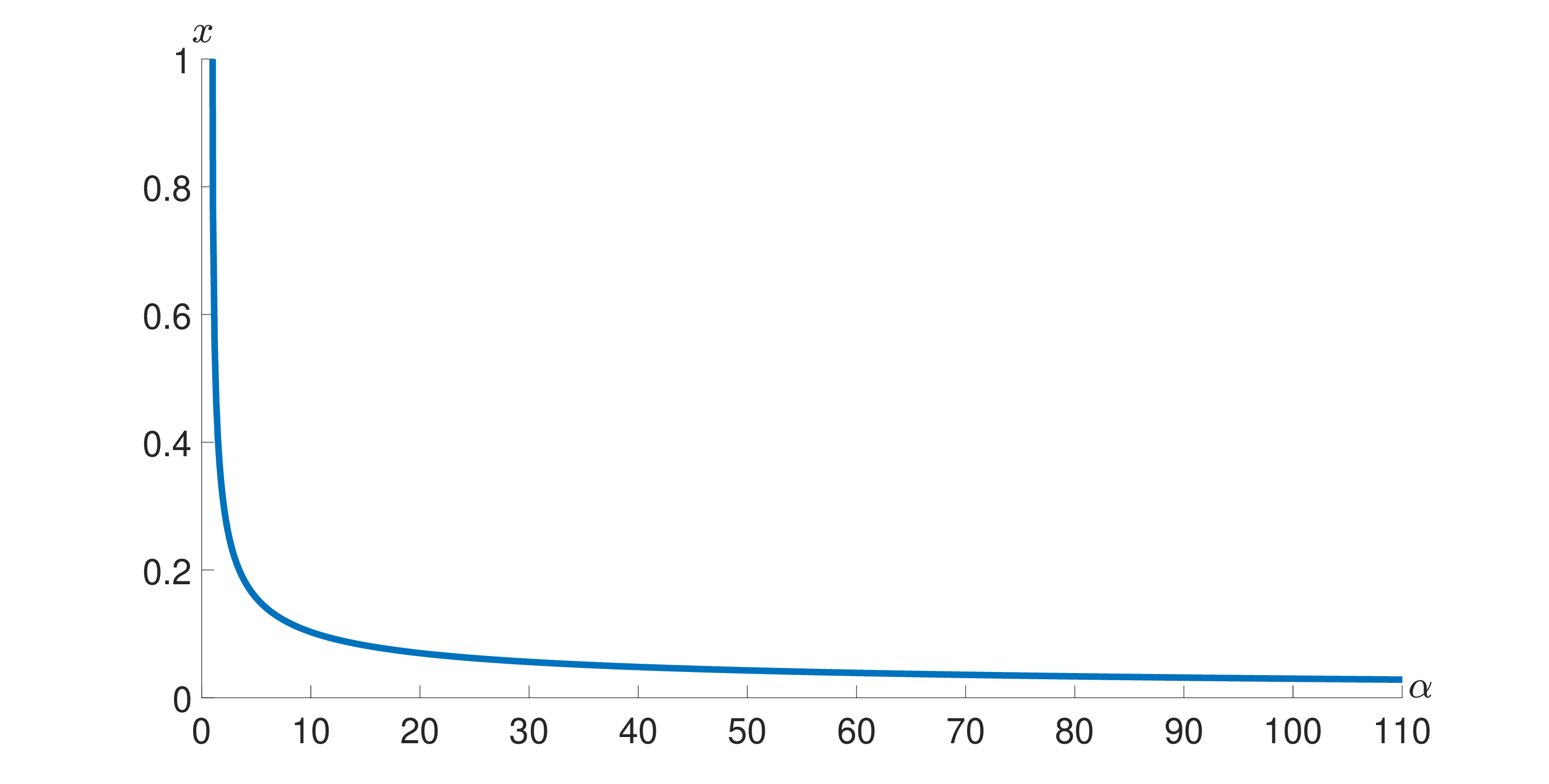}
\vskip -4mm \caption{The ${x_{\it{min}}}-\alpha$ curve }\label{fig3.1}
\end{figure}

As illustrated in Fig. \ref{fig3.1}, it can be observed that $x_{\it{min}}$ towards zero with an increase in $\alpha$. This is demonstrated by the values of $x_{\it{min}}$, which are 0.0560513957; 0.042749089; 0.029789688, respectively, when the values of $\alpha$ are 30; 50; 100, respectively. It can be concluded that the position ratio $x$ between the two horizons assumes a value within the interval $x_{\it{min}}\leq{x}\leq1$ when the coexistence region of the two horizons exists in spacetime.

When the potential on the black hole horizon in spacetime satisfies the following expression \cite{Mellor1989}
\begin{align}\label{3.10}
  \frac{Q^2}{r_{+}^{2}} & =\frac{1}{(1+x)^2},
\end{align}
the radiation temperature of the black hole horizon, $\bar{T}_{+}$, is equal to that of the cosmological horizon, $\bar{T}_{c}$, i.e.
\begin{align}\label{3.11}
  \bar{T}_{+} & =\bar{T}_{c}=\frac{1-x}{2{\pi}r_c(1+x)^2}.
\end{align}
Substituting Eq. (\ref{3.10}) into Eq. (\ref{3.5}), we obtain
\begin{align}\label{3.12}
T_{\it{eff}} & =\bar{T}_{\it{eff}}=\bar{T}_{\it{+}}+\frac{(1-x)}{2{\pi}r_{c}x^4(1+x)^2}=\bar{T}_c+\frac{(1-x)}{2{\pi}r_{c}x^4(1+x)^2}.
\end{align}
From Eq. (\ref{3.12}), it can be observed that treating the thermodynamic system corresponding to the two horizons with the same radiation temperature in RN-dS spacetime as an independent thermodynamic system leads to a discrepancy in the effective radiation temperature which takes into account the conditions of interaction between the two horizons. It can be demonstrated that the effective temperature, $\bar{T}_{\it{eff}}$, of an effective thermodynamic system consisting of two subsystems with the same temperature in RN-dS spacetime, is not equal to the temperature of the subsystems, i.e. $\bar{T}_{\it{eff}}>\bar{T}_+=\bar{T}_c$. This phenomenon can be attributed to the interaction between the two horizons. This phenomenon can be attributed to the existence of an interaction between the two subsystems. Despite the temperatures of the two subsystems being identical, namely, $\bar{T}_+=\bar{T}_c$, the interaction between them results the effective temperature, $\bar{T}_{\it{eff}}$, of the effective thermodynamic system being distinct from that of the respective subsystems. This illustrates the pivotal role of the interaction between the two horizons in RN-dS spacetime in the effective thermodynamic system. This demonstrates that the thermodynamic system in the RN-dS spacetime is distinct from the ordinary thermodynamic system. The discovery of this phenomenon suggests that it is necessary to take into account the behavior of thermodynamic quantities due to the interaction between black holes and the external environment when the thermodynamic properties of black holes are studied by analogy with the method used to study ordinary thermodynamic systems. Consequently, it is of the importance to investigate the intrinsic causes of the interaction between the two horizons in greater depth.

The behavior of $T_+$, $T_c$ and $T_{\it{eff}}$ can be plotted against the ratio between the two horizon positions, $x$, for different values of the parameter $\alpha$ by substituting Eq. (\ref{2.12}) into Eqs. (\ref{2.13}), (\ref{2.14}) and (\ref{3.5}) (taking a $+$ sign in Eq. (\ref{2.12})).
\begin{figure}[htb]
\centering
\includegraphics[width=8.5cm,height=4.5cm]{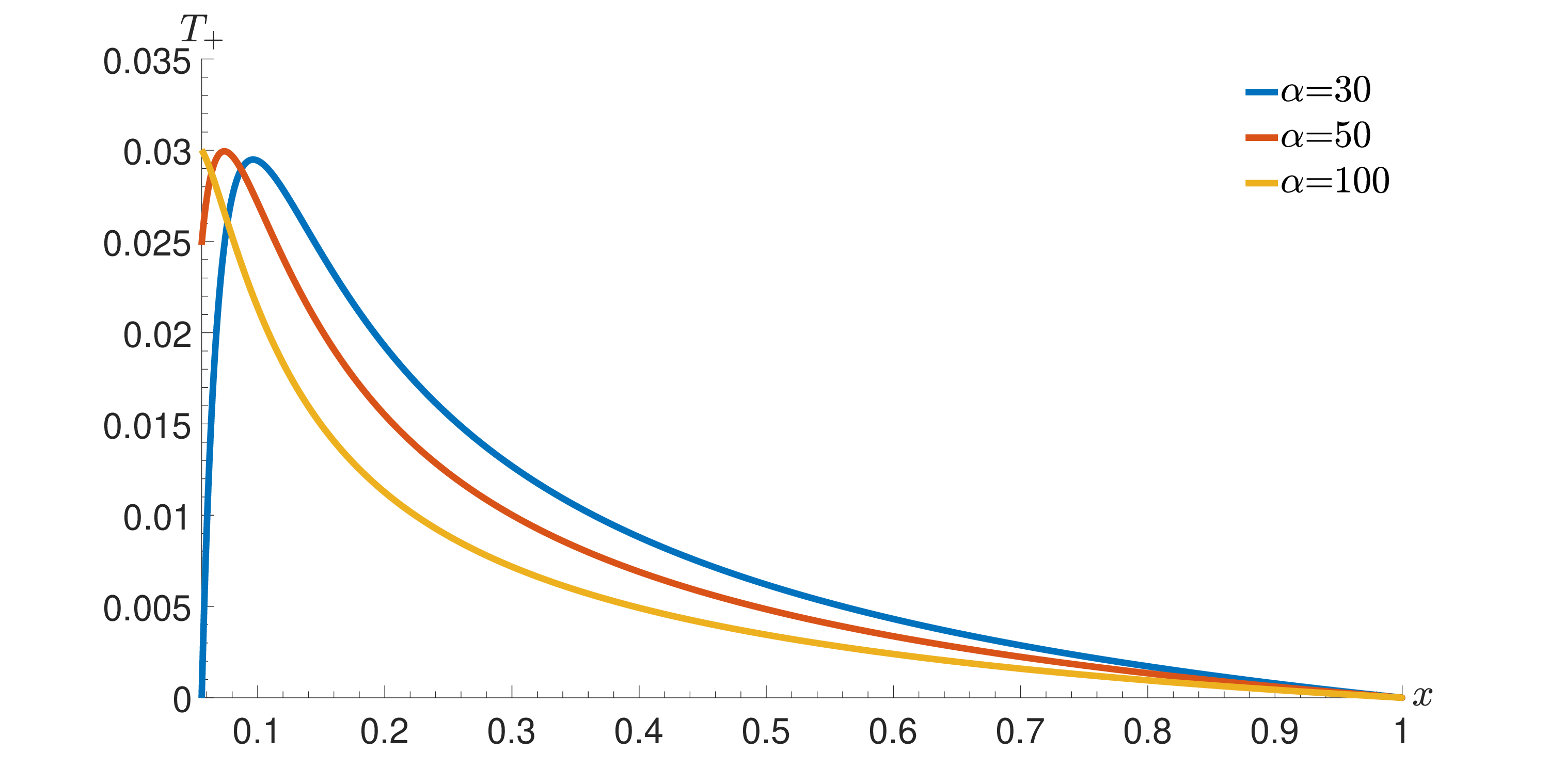}
\vskip -4mm \caption{The ${T_{+}}-x$ curve for different $\alpha$, $Q$=1}\label{fig3.2}
\end{figure}
\begin{figure}[htb]
\centering
\includegraphics[width=8.5cm,height=4.5cm]{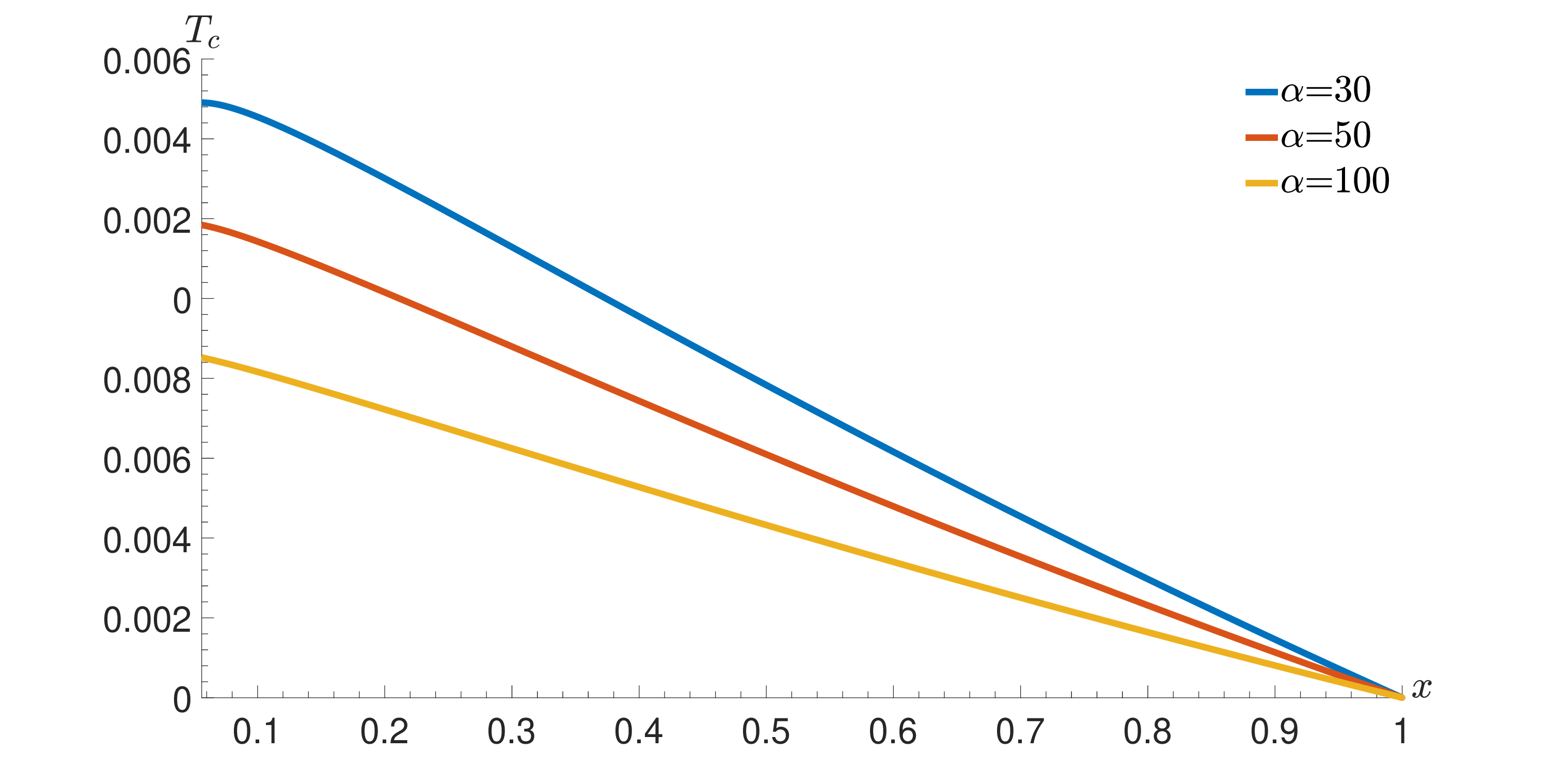}
\vskip -4mm \caption{The ${T_{c}}-x$ curve for different $\alpha$, $Q$=1}\label{fig3.3}
\end{figure}
\begin{figure}[htb]
\centering
\includegraphics[width=8.5cm,height=4.5cm]{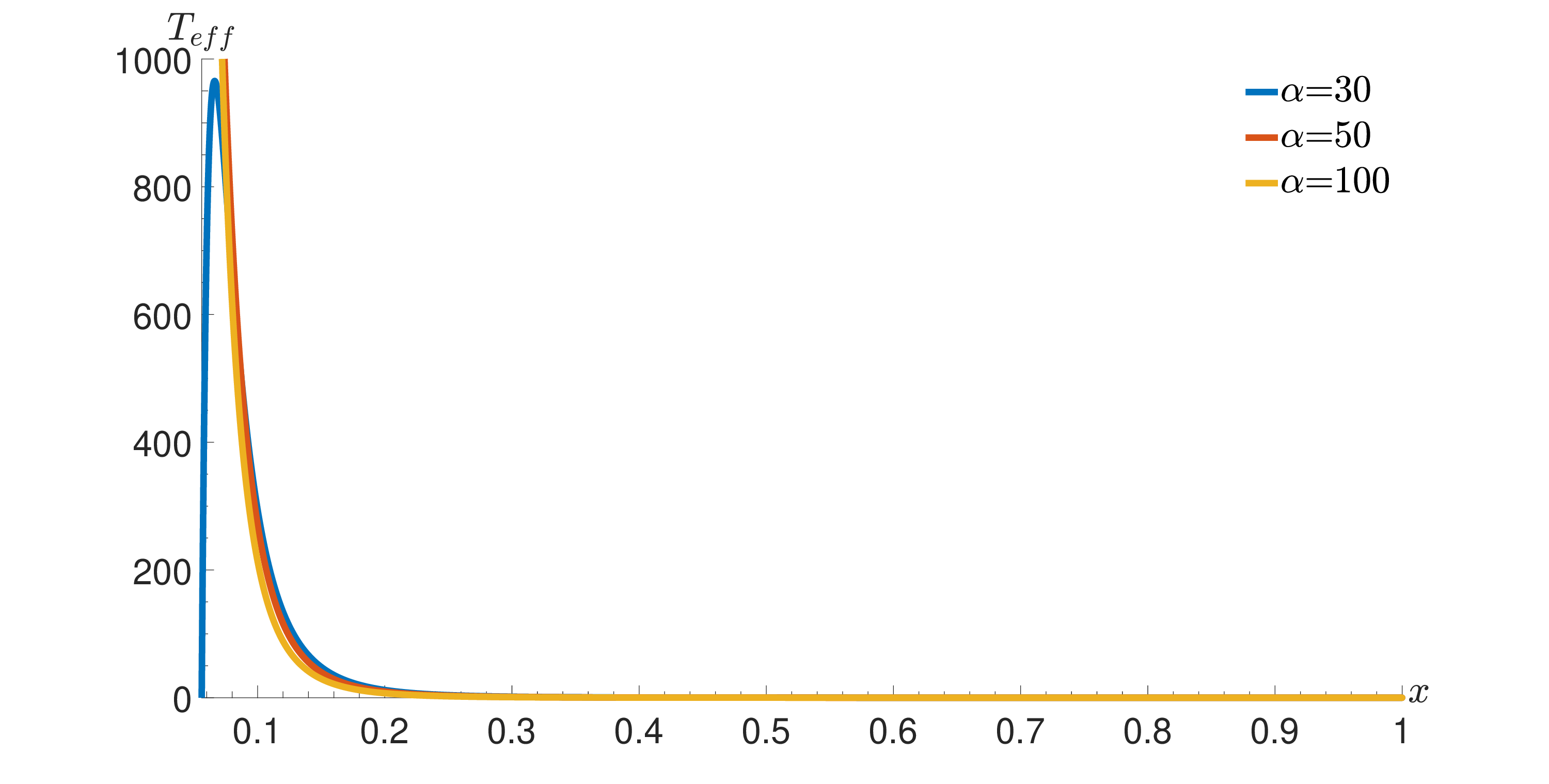}
\vskip -4mm \caption{The ${T_{\it{eff}}}-x$ curve for different $\alpha$, $Q$=1}\label{fig3.4}
\end{figure}

\begin{figure}[htb]
\centering
\includegraphics[width=8.5cm,height=4.5cm]{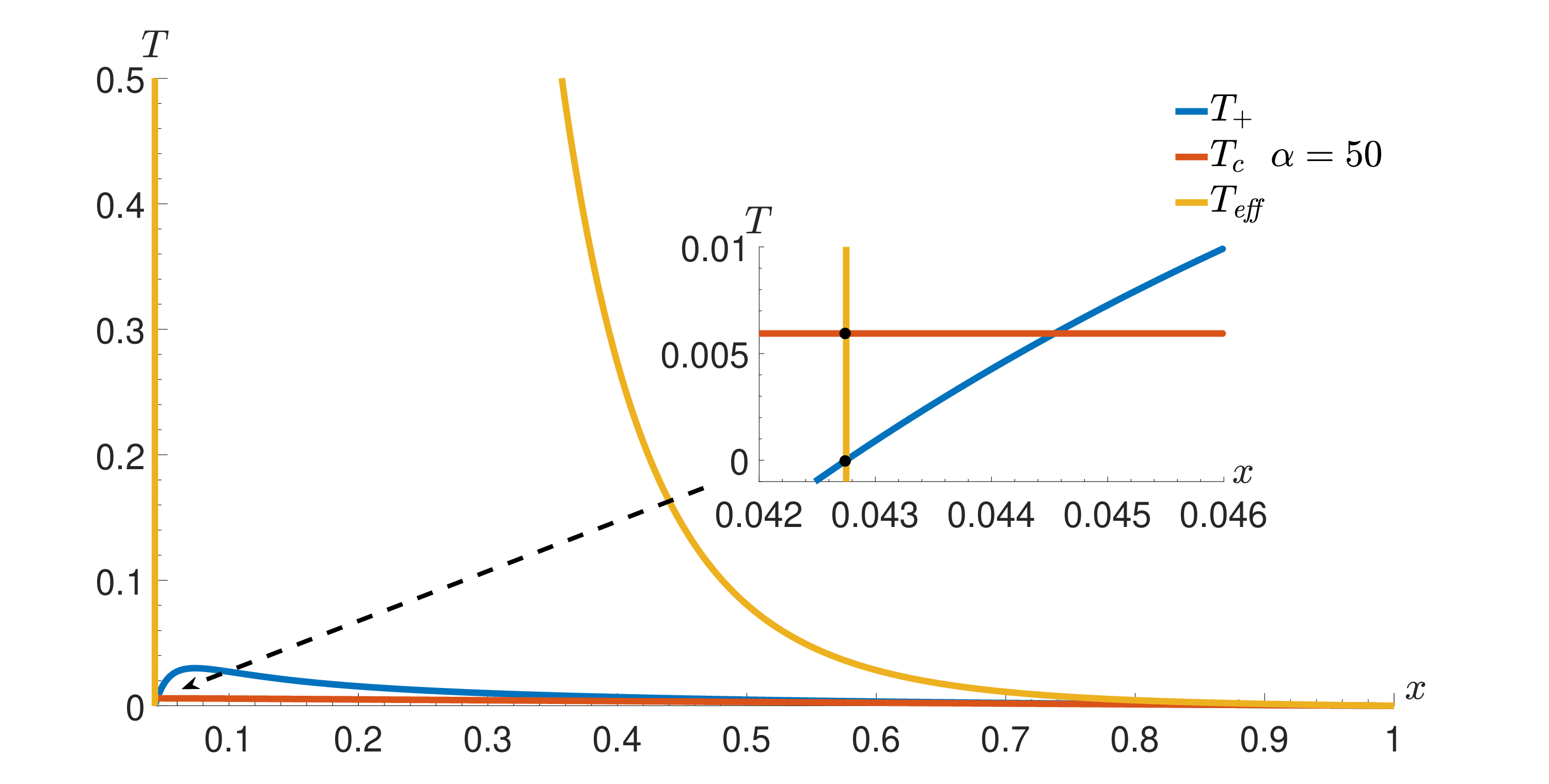}
\vskip -4mm \caption{The ${T_{+},T_{c},T_{\it{eff}}}-x$ curves for $\alpha$=50, $Q$=1}\label{fig3.5}
\end{figure}

Figs. \ref{fig3.2}, \ref{fig3.3} and \ref{fig3.4} illustrate that $T_+$, $T_c$ and $T_{\it{eff}}$ decrease as $\alpha$ increases at a fixed value of $x$. The behavior of $T_+$, $T_c$ and $T_{\it{eff}}$ as a function of $x$ is plotted in Fig. \ref{fig3.5}

Fig. \ref{fig3.5} demonstrates that the temperature in question,which is a consequence of the interaction between the two horizons, is constrained to satisfy $T_{\it{eff}}\geq{T_+}\geq{T_c}$ under a fixed $x$. The boundary condition of the effective thermodynamic system is satisfied when the effective temperature, $T_{\it{eff}}$, and the radiation temperature of the cosmological horizon, $T_c$, are equal, as well as the radiation temperature of the black hole horizon, $T_+$, is zero. And in a small region of the smaller ratio between two horizons, the black hole temperature increases with the the increasing of $x$, while the cosmological one keeps nearly a constant, until they equal to each other, this is maybe from the interplay between two horizons. Then the black hole temperature is always high than the cosmological temperature. Furthermore, the Smarr relation of the effective thermodynamic system is as follows:
\begin{align}\label{3.13}
M&=2T_{\it{eff}}S+{\varphi}_{\it{eff}}Q-3P_{\it{eff}}V
\end{align}
\section{Schottky anomaly of RN-dS black holes }\label{four}

\begin{figure}[htb] {\includegraphics[width=8.8cm,height=4.5cm]{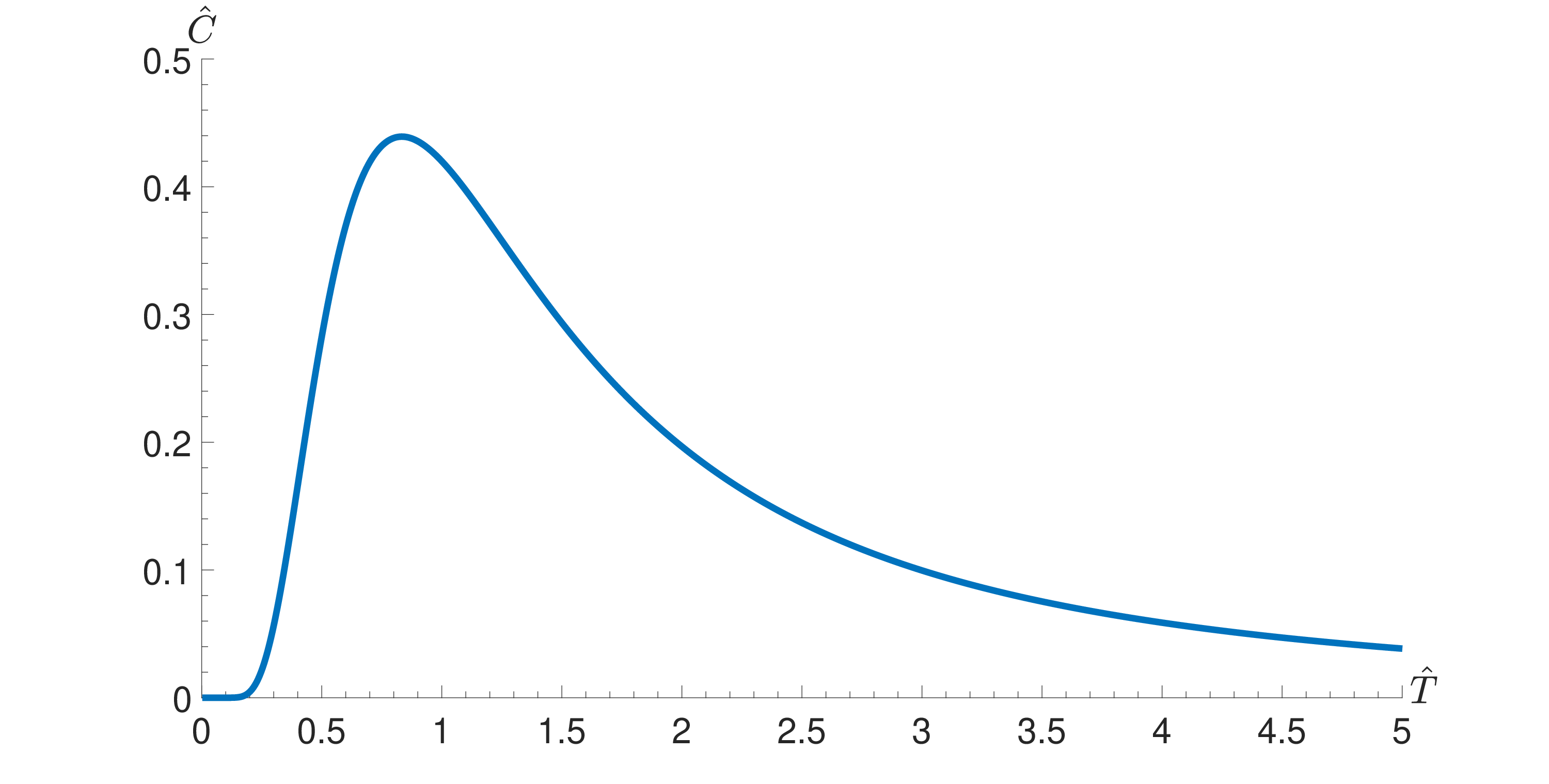}\label{fig4.1b}}~
\vskip -4mm \caption{The $\hat{C}-\hat{T}$ curve with $\hat{T}=T/\varepsilon$. }\label{fig4.1}
\end{figure}

The heat capacity of a thermal system is typically an increasing function of temperature. However, a peak in the heat capacity, known as a Schottky anomaly, can occur in a system that has a maximum energy. It is well established that for a paramagnetic mass system with $j$ = 1/2, which is a two-level system, the partition function reads
\begin{align}\label{4.1}
  z & =e^{{\beta}{\varepsilon}}+e^{-{\beta}{\varepsilon}},
\end{align}
where $\beta=1/kT$, $\pm\varepsilon$ stand for the energy of two energy levels. From Eq. (\ref{4.1}), the internal energy of the system has the form as follows:
\begin{align}\label{4.2}
  U & =F+TS=-N{\varepsilon}\tanh({\beta}{\varepsilon}).
\end{align}
Neglecting the interaction between particles the specific heat can be expressed in the following form
\begin{align}\label{4.3}
  \hat{C} & =Nk\left(\frac{\triangle}{T}\right)^2\frac{e^{\frac{\triangle}{T}}}{(1+e^{\frac{\triangle}{T}})^2}.
\end{align}
This is a characteristic of a two-energy-level system with an energy gap $\Delta=2{\varepsilon}$, which is just the celebrated Schottky specific heat. It is of great importance for the treatment of a multitude of systems as two-level systems. The behaviour of heat capacity as the function of temperature is presented in Fig. \ref{fig4.1} (see more details in refs. \cite{Chakrabhavi2311,Dinsmore2020}.)

The effective thermodynamic system, comprising with two thermodynamic subsystems, was obtained through the consideration of the interaction between the two horizons with disparate temperatures in dS spacetime. The interaction between the two subsystems occurs through the two horizons. It is therefore necessary to determine whether the thermodynamic properties between the two horizons can be described as those of a two-level system. This is one of the key issues to be addressed in the study of the thermodynamic properties in dS spacetime. In recent years, the Schottky specific heat with some promising results emerging in dS spacetime has been investigated \cite{Chakrabhavi2311,Dinsmore2020,Johnson2020}. Nevertheless, the conclusions obtained are not exhaustive in the absence of the interaction between the black hole and cosmological horizons. The Schottky specific heat in RN-dS spacetime based on the interaction between the two horizons is investigated in this section.

When the charge, $Q$, and the cosmological constant, $l$, in spacetime are held constant, based on these effective thermodynamical quantities the heat capacity of this RN-dS spacetime can be obtained by the following expression
\begin{align}\label{4.4}
  C_{Q,l} & =T_{\it{eff}}\left(\frac{\partial{S}}{\partial{T_{\it{eff}}}}\right)_{Q,l}=T_{\it{eff}}\left(\frac{dS/dx}{dT_{\it{eff}}/dx}\right)_{Q,l}.
\end{align}
From Eq. (\ref{4.4}) the behavior of $C_{Q,l}$ with respect to the ratio between two horizons and the effective temperature can be plotted under different values of $\alpha$.

\begin{figure}[htb]
\centering
\includegraphics[width=8.5cm,height=4.5cm]{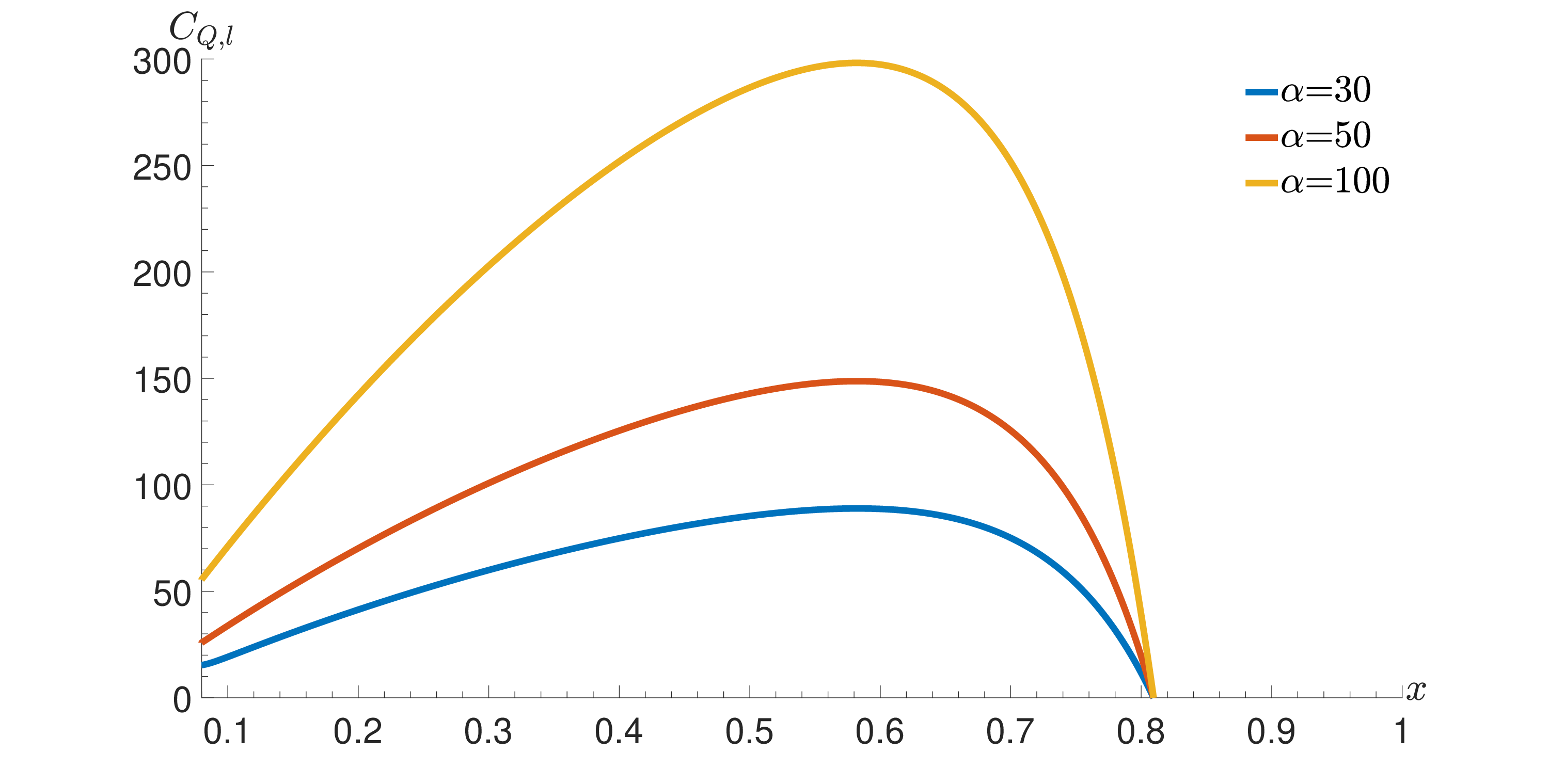}
\vskip -4mm \caption{The ${C_{\it{Q,l}}-x}$ curve for different $\alpha$}\label{fig4.2}
\end{figure}

\begin{figure}[htb]
\centering
\includegraphics[width=8.5cm,height=4.5cm]{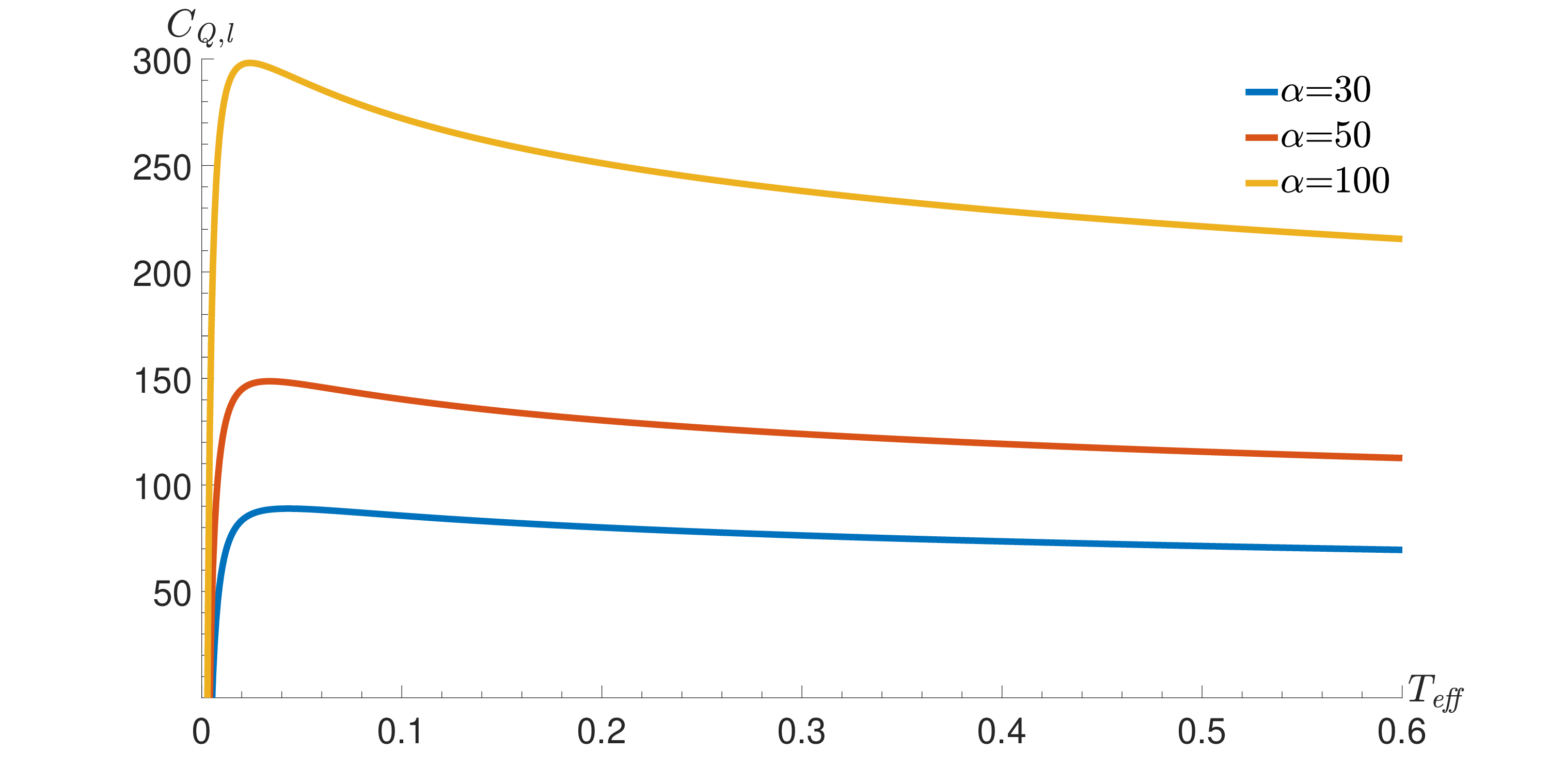}
\vskip -4mm \caption{The ${C_{\it{Q,l}}-T_{\it{eff}}}$ curve for different $\alpha$ }\label{fig4.3}
\end{figure}

\begin{figure}[htb]
\centering
\includegraphics[width=8.5cm,height=4.5cm]{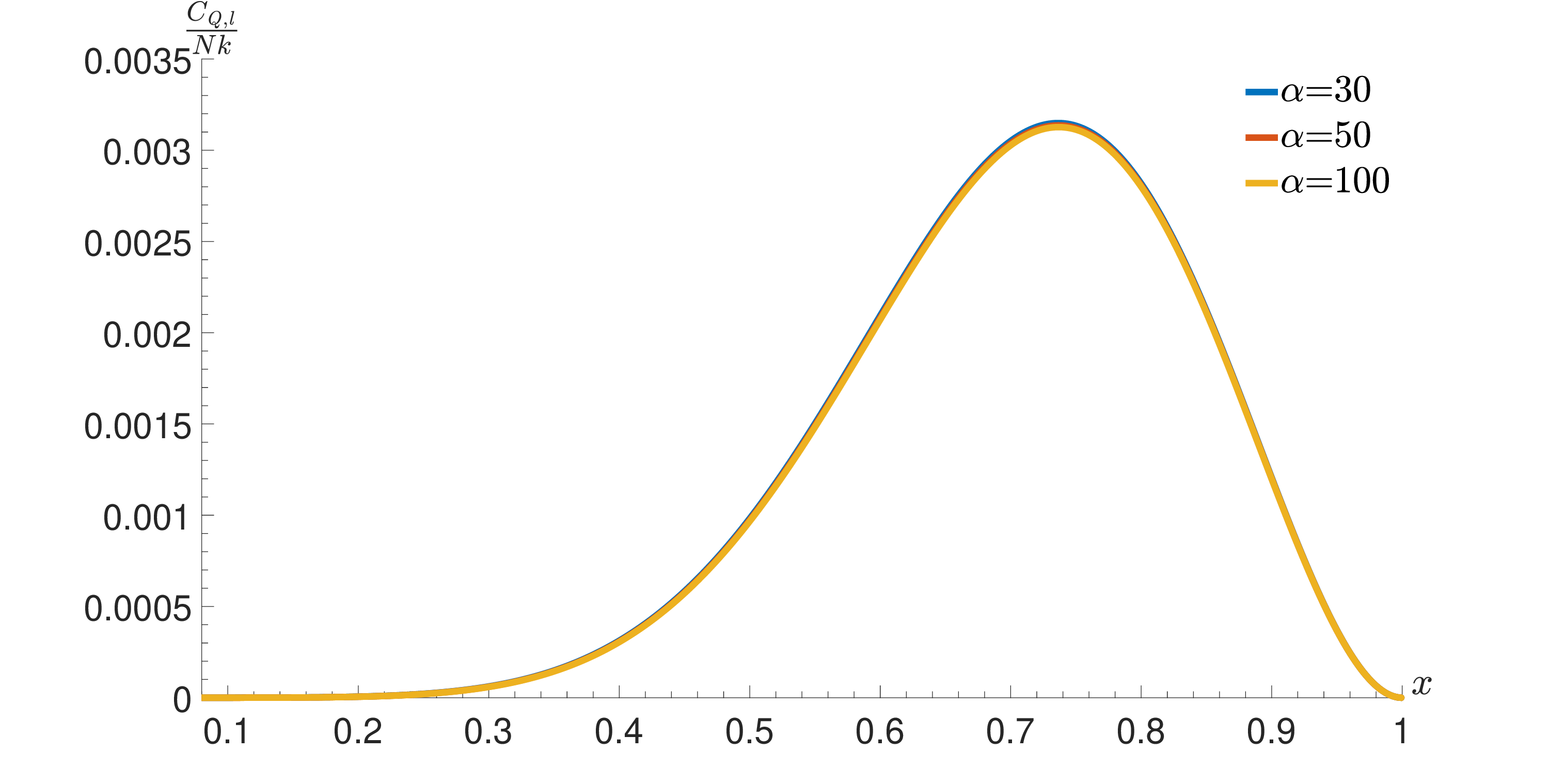}
\vskip -4mm \caption{The $\hat{C}_{\it{Q,l}}/{Nk}-x$ curve for different $\alpha$, $\varepsilon$=1 }\label{fig4.6}
\end{figure}

\begin{figure}[htb]
\centering
\includegraphics[width=8.5cm,height=4.5cm]{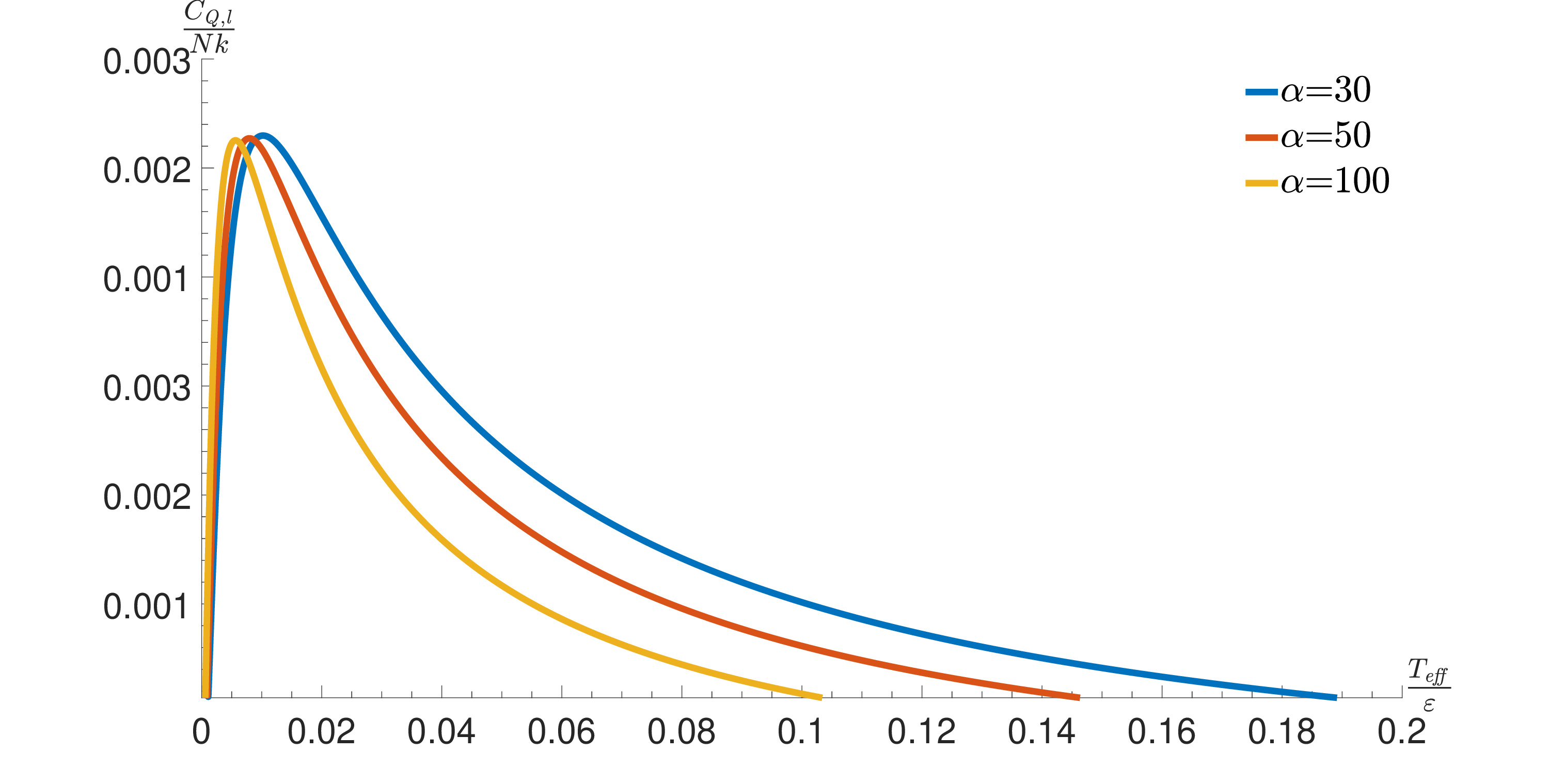}
\vskip -4mm \caption{The $\hat{C}_{\it{Q,l}}/{Nk}-T_{\it{eff}}/\varepsilon$ curve for different $\alpha$ }\label{fig4.7}
\end{figure}

As illustrated in Figs. \ref{fig4.2} and \ref{fig4.3}, the heat capacity of RN-dS spacetime described by the effective quantities exhibits similarities to that of the two-level system shown in Fig. \ref{fig4.1}, namely both they exhibits a pattern behavior. The heat capacity has an extreme value within certain parameter ranges. Note that when the space between two horizons in RN-dS spacetime is depicted by the corresponding effective quantities we call it as the effective thermodynamic system.
The effective thermodynamic system in RN-dS spacetime is constituted by the subsystems that correspond to the two horizons. Consequently, the two horizons in RN-dS spacetime can be regarded as two distinct energy levels of a two-level system. When the particles are situated on different horizons, they exhibit distinct energy levels. According to this hypothesis, the two horizons in RN-dS spacetime are regarded as two different energy levels. Thus the energies of the particles on the black hole and cosmological horizons read
\begin{align}\label{4.5}
  {\varepsilon}_{+}&={\varepsilon}T_{+},~~{\varepsilon}_{c}={\varepsilon}T_{c},
\end{align}
where $\varepsilon$ is a positive constant. And the term of $\Delta$ in Eq. (\ref{4.3}) becomes
\begin{align}\label{4.6}
  \Delta & ={\varepsilon}_{+}-{\varepsilon}_{c}={\varepsilon}\left(T_{+}-T_{c}\right),~~T=T_{\it{eff}}/{\varepsilon}.
\end{align}
Since the behavior of heat capacity for the RN-dS spacetime regarded as a two-level system is independent of the parameter $\varepsilon$, so we set it to $\varepsilon=1$. Substituting Eqs. (\ref{2.13}), (\ref{2.14}), and (\ref{3.5}) into Eq. (\ref{4.6}) we can obtain the following form
\begin{align}\label{4.7}
  \frac{\Delta}{T_{\it{eff}}} & =\frac{{\varepsilon}x^4\left(1-x\right)^2\left(1+x\right)
  \left(1-\frac{1}{2}\left(1-\sqrt{1-\frac{\left(1+x+x^2\right)}
  {3x\alpha}}\right)\frac{\left(1+x\right)^2}{x}\right)}
  {\left(1-x^3\right)\left\{\left[\left(1+x\right)
  \left(1+x^3\right)-2x^2\right]-\frac{1}{2x}
  \left(1-\sqrt{1-\frac{\left(1+x+x^2\right)}{3x\alpha}}\right)
  \left[\left(1+x+x^2\right)\left(1+x^4\right)-2x^3\right]\right\}}.
\end{align}
In order to differentiate it from a general two-level system, we denote the heat capacity of RN-dS spacetime as a two-level system by the following form
\begin{align}\label{4.8}
  \hat{C}_{Q,l} & =Nk\left(\frac{{\varepsilon}{\Delta}}{T_{\it{eff}}}\right)^2
  \frac{e^{\frac{{\varepsilon}{\Delta}}{T_{\it{eff}}}}}
  {\left(1+e^{\frac{{\varepsilon}{\Delta}}{T_{\it{eff}}}}\right)^2}.
\end{align}
The corresponding behavior respect to the ratio between two horizons and the effective temperature are shown in Figs. \ref{fig4.6} and \ref{fig4.7}.

A comparison of Figs. \ref{fig4.2} and \ref{fig4.6}, and Figs. \ref{fig4.3} and \ref{fig4.7}, reveals that two horizons in the RN-dS spacetime are indeed two distinct energy levels of the effective thermodynamic system. Therefore the RN-dS spacetime is a quantum system with two separate energy levels. And its heat capacity $\hat{C}_{Q,l}$ approaches to zero when the temperature tends to zero. With the increasing of temperature $\hat{C}_{Q,l}$ reaches to a maximum and then decreases to zero, which is the famous Schottky behavior. It is significant that numerous systems can be treated as a two-level system. This result demonstrates a novel quantum property in the RN-dS spacetime.

\section{Conclusion}
\label{five}

In this manuscript we considered a new boundary condition of RN-dS spacetime with the interplay between the black hole and cosmological horizons and gave the corresponding thermodynamic effective quantities. As well as the Smarr relation and the system free energy expressed in the effective thermodynamic quantities were presented. These are the foundation for further study of the thermodynamic properties in RN-dS spacetime. By analyzing the heat capacity of RN-dS spacetime with the coexistence of two horizons, we found the heat capacity of RN-dS spacetime described by the effective thermodynamic quantities has the similar behavior of the Schottky specific heat. In order to elucidate the underlying cause of this phenomenon, we proposed that the black hole and cosmological horizons can be regarded as two distinct energy levels of the RN-dS spacetime which is an effective thermodynamic system, i.e. a two-level system. The heat capacity of this two-level system was discussed by a comparison between Figs. \ref{fig4.2} and \ref{fig4.6}, and Figs. \ref{fig4.3} and \ref{fig4.7}. It was found that the behavior of the heat capacity of this two-level system is highly similar to that of the effective thermodynamic system. Consequently, it can be concluded that the behavior of the heat capacity, $C_{Q,l}$, in the RN-dS spacetime is primarily determined by the heat capacity, $\hat{C}_{Q,l}$, of the two-level system.

In addition, the conclusions showed that the RN-dS spacetime with two separate energy levels is a quantum system, and when the effective temperature tends to zero the corresponding heat capacity $\hat{C}$ approaches to zero. The heat capacity $\hat{C}$ also has an extreme value, and is in accordance with that in a two-level system, i.e. the well-known Schottky specific heat. Furthermore, we had investigated the effect of different parameters in RN-dS spacetime on the behavior of heat capacity. A comprehensive investigation of this topic would provide a deeper understanding of the quantum properties of black holes. At the same time, it would also suggest a new avenue for research into the thermodynamic properties of dS spacetime. This result encapsulates the quantum properties of RN-dS spacetime and offers a novel avenue for further in-depth investigation into the quantum properties of black holes and dS spacetime. In future we will discuss whether the behavior of heat capacity of thermodynamic systems in RN-dS spacetime is universal to that in other different dS spacetime with black holes, and whether the parameters in other different dS spacetime can affect the property of heat capacity.

\section*{Acknowledgments}
We would like to thank Prof. Ren Zhao and Meng-Sen Ma for their indispensable discussions and comments. This work was supported by the Natural Science Foundation of China (Grant No. 12375050, Grant No. 11705106, Grant No. 12075143), the Scientific Innovation Foundation of the Higher Education Institutions of Shanxi Province (Grant Nos. 2023L269), the Science Foundation of Shanxi Datong University (2022Q1, 2020Q2), the Natural Science Foundation of Shanxi
Province (Grant No. 202203021221209, Grant No. 202303021211180) and the Teaching Reform Project of Shanxi Datong University (XJG2022234).


\begin{references}

\bibitem{M871} The event horizon telescope collaboration et al, ApJL 875 L1 (2019)
\bibitem{M872} The event horizon telescope collaboration et al, ApJL 875 L2 (2019)
\bibitem{HSW} Hawking S. W, Commun. Math. Phys, 43(3): 199-220 (1975)
\bibitem{BJD} Bekenstein J. D, Phys. Rev. D, 7: 2333-2346 (1973)
\bibitem {Kubiznak2012} D. Kubiznak, R. B. Mann, JHEP, 1207: 033 (2012)
\bibitem{Cai2013} R.G. Cai, L.M. Cao, L. Li et al, JHEP, 09: 005 (2013)
\bibitem{Cai2016} R.G. Cai, S.M. Ruan, S.J. Wang et al, JHEP, 09: 161 (2016)
\bibitem{JLZhang2015} J.L. Zhang, R.G. Cai, H.W. Yu, Phys. Rev. D, 91: 044028 (2015)
\bibitem{Gunasekaran2012} S. Gunasekaran, D. Kubiznak, R. B. Mann, JHEP, 11: 110 (2012)
\bibitem{Frassino2014} A. M. Frassino, D. Kubiznak, R. B. Mann et al, JHEP, 09: 080 (2014)
\bibitem{Altamirano2014} N. Altamirano, D. Kubiznak, R. B. Mann et al, Class. Quant. Grav, 31: 042001 (2014)
\bibitem{JYYang2023} J.Y. Yang, R. B. Mann, JHEP, 08: 028 (2023)
\bibitem{JWu2023} J. Wu, R. B. Mann, Phys. Rev. D, 107: 084035 (2023)
\bibitem{SWWei2303} S.W. Wei , Y.P. Zhang, Y.X. Liu et al, Phys. Rev. Res, 5: 043050 (2023)
\bibitem{Durgut2023} T. Durgut, R. A. Hennigar, H. K. Kunduri et al, JHEP, 03: 114 (2023)
\bibitem{JWu20231} J. Wu, R. B. Mann, Class. Quan. Grav, 40: 06LT01 (2023)
\bibitem{YXiao2024} Y. Xiao, Y. Tian, Y.X. Liu, Phys. Rev. L, 132: 021401 (2024)
\bibitem{MTW2023} M.T. Wang, H.Y. Jiang, Y.X. Liu, JHEP 07, 178 (2023)
\bibitem{RZhao2013} R. Zhao, H.H. Zhao, M.S. Ma et al, Eur. Phy. J. C, 73: 2645 (2013)
\bibitem{Hendi2017} S. H. Hendi, R. B. Mann, S. Panahiyan et al, Phys. Rev. D, 95: 021501(R) (2017)
\bibitem{Hendi2023} S. H. Hendi, S. Hajkhalili, M. Jamil et al, Fort. der. Phy, 71(8-9) (2023)
\bibitem{Momennia2021} M. Momennia, S. H. Hendi, Phys. Lett. B, 822: 136692 (2021)
\bibitem{Hendi2021} S. H. Hendi, K. Jafarzade, Phys. Rev. D, 103: 104011 (2021)
\bibitem{Laassiri2401} H. Laassiri, A. Daassou, R. Benbrik, Int. J. Mod. Phys. A, 39(24): 2450089 (2024)
\bibitem{XQLi2023} X.Q. Li, H.P. Yan, L.L. Xing et al, Phys. Rev. D, 107: 104055 (2023)
\bibitem{Kruglov2022} S. I. Kruglov, Eur. Phys. J. C, 82(4): 292 (2022)
\bibitem{Kruglov20221} S. I. Kruglov, Annals Phys., 441: 168894 (2022)
\bibitem{YZDu2022} Y.Z. Du, H.F. Li, Ren Zhao, Eur. Phys. J. C, 82: 859 (2022)
\bibitem{YZDu2023} Y.Z. Du, H.F. Li, F. Liu et al, JHEP, 01: 137 (2023)
\bibitem{ZYGao2022} Z.Y. Gao, X.Q. Kong, L. Zhao, Eur. Phys. J. C, 82: 112 (2022)
\bibitem{Ghaffari2312} S. Ghaffari, G. G. Luciano, A. Sheykhi, Phys. Dark Univ., 44: 101447 (2024)
\bibitem{MYZhang2401} M.Y. Zhang, H. Chen, H. Hassanabadi et al, Chin. Phys. C, 48(6): 065101 (2024)
\bibitem{DJGogoi2023} D. J. Gogoi, Y. Sekhmani, D. Kalita et al, Fort. Der. Phy, 71: 2300010 (2023)
\bibitem{Yasir2024} M. Yasir, X. Tiecheng, F. Javed et al, Chin. Phys. C, 48: 015103 (2024)
\bibitem{Barrientos2022} J. Barrientos, J. Mena, Phys. Rev. D, 106: 044064 (2022)
\bibitem{Sadeghi2402} J. Sadeghi, M. R. Alipour, S. N. Gashti et al, arXiv:2402.02257 [hep-th]
\bibitem{Hunga2023} T. N. Hunga, C. H. Nam, Eur. Phys. J. C, 83: 582 (2023)
\bibitem{Gogoi2023} N. J. Gogoi, P. Phukon, Phys. Rev. D, 108: 066016 (2023)
\bibitem{Fairoos2304} C. Fairoos, and T. Sharqui, Int. J. Mod. Phys. A, 38: 2350133 (2023)
\bibitem{DWu2402} D. Wu, S.Y. Gu, X.D. Zhu et al, JHEP, 06: 213 (2024)
\bibitem{Yerra2304} P. K. Yerra,  C. Bhamidipati, S. Mukherji, JHEP, 03: 138 (2024)
\bibitem{SWWei2015} S.W. Wei, Y.X. Liu, Phys. Rev. D, 91: 044018 (2015)
\bibitem{SWWei20151} S.W. Wei, Y.X. Liu, Phys. Rev. Lett., 115: 111302 (2015)
\bibitem{SWWei2019} S.W. Wei, Y.X. Liu, R. B. Mann, Phys. Rev. Lett, 123: 071103 (2019)
\bibitem{DYChen2023} D.Y. Chen, J. Tao, X.T. Yang, Phys. Dark. Universe, 42: 101379 (2023)
\bibitem{Sokoliuk2024} O. Sokoliuk, S. Pradhan, A. Baransky et al, Fort. Phys, 72: 2300043 (2024)
\bibitem{Ruppeiner2023} G. Ruppeiner, A. M. Sturzu, Physical Review D, 108: 086004 (2023)
\bibitem{Masood2023} S. Masood A. S. Bukhari, B. Pourhassan, H. Aounallah et al, Class. Quant. Grav, 40: 225007 (2023)
\bibitem{GRLi2022} G.R. Li, G.P. Li, S. Guo, Class. Quant. Grav, 39: 195011 (2022)
\bibitem{XYGuo2019} X.Y. Guo, H.F. Li, L.C. Zhang et al, Phys. Rev. D, 100: 064036 (2019)
\bibitem{RGCai2002} R.G. Cai, Nucl. Phys. B, 628: 375-386 (2002)
\bibitem{Marks2107} G. A. Marks, F. Simovic, R. B. Mann, Phys. Rev. D, 104: 104056 (2021)
\bibitem{Sekiwa2006} Y. Sekiwa, Phys. Rev. D, 73: 084009 (2006)
\bibitem{Urano2009} M. Urano, A. Tomimatsu, H. Saida, Class. Quant. Grav, 26: 105010 (2009)
\bibitem{Mbarek2019} S. Mbarek, R. B. Mann, JHEP, 02: 103 (2019)
\bibitem{Kubiznak2016} D. Kubiznak,  F. Simovic, Class. Quan. Grav, 33: 245001 (2016)
\bibitem{Simovic2019} F. Simovic, R. B. Mann, JHEP, 05: 136 (2019)
\bibitem{Haroon2020} S. Haroon, R. A. Hennigar, R. B. Mann et al, Phys. Rev. D, 101: 084051 (2020)
\bibitem{Simovic2008} F. Simovic, D. Fusco, R. B. Mann, JHEP, 02: 219 (2021)
\bibitem{Dolan2013} B. P. Dolan, D. Kastor, D. Kubiznak et al, Phys. Rev. D., 87: 104017 (2013)
\bibitem{Bhattacharya2016} S. Bhattacharya, Eur. Phys. J. C, 76: 112 (2016), [arXiv:1506.07809]
\bibitem{McInerney2016} J. McInerney, G. Satishchandrana, J. Traschena, Class. Quantum Grav, 33: 105007 (2016)
\bibitem{Kanti2017} P. Kanti, T. Pappas, Phys. Rev. D, 96: 024038 (2017)
\bibitem{LCZhang2016} L.C. Zhang, R. Zhao, M.S. Ma, Phys. Lett. B, 761: 74-76 (2016)
\bibitem{LCZhang2019} L.C. Zhang, R. Zhao, Phys. Lett. B, 797: 134798 (2019)
\bibitem{YBMa2020} Y.B. Ma, Y. Zhang, L.C. Zhang et al, Eur. Phys. J. C, 80: 213 (2020)
\bibitem{Ko2312} J. Ko, B. Gwak, JHEP, 03: 072 (2024)
\bibitem{Chakrabhavi2311} V. Chakrabhavi, M. Etheredge, Y. Qiu et al, JHEP, 02: 231 (2024)
\bibitem{Anderson2022} P. R. Anderson, J. Traschen, JHEP, 01: 192 (2022)
\bibitem{Dinsmore2020} J. Dinsmore, P. Draper, D. Kastor et al, Class. Quan. Grav, 37: 054001 (2020)
\bibitem{YBDu2303} Y.B. Du, X.D. Zhang, Eur. Phys. J. C, 83: 927 (2023)
\bibitem{Nam2018} C. H. Nam, Eur. Phys. J. C, 78: 418 (2018)
\bibitem{Mellor1989} F. Mellor, I. Moss, Class. Quan. Grav, 6: 1379 (1989)
\bibitem{Johnson2020} C. V. Johnson, Class. Quant. Grav, 37: 054003 (2020)




\end{references}
\end{document}